\documentclass[prl,aps,twocolumn,showpacs]{revtex4}

\usepackage{graphicx}
\newcommand{\upp}{\hspace{-3 pt}\uparrow}
\newcommand{\downn}{\hspace{-3 pt}\downarrow}

\begin{document}

\bibliographystyle{apsrev}

\title{Long-lived memory for mesoscopic quantum bits}
\author{J.M. Taylor}
\author{C.M. Marcus}
\author{M.D. Lukin }
\affiliation{Department of Physics, Harvard University, Cambridge,
Massachusetts 02138}

\date{\today}
\begin{abstract}
We describe a technique to create long-lived quantum memory for quantum 
bits in mesoscopic systems. Specifically  we show that electronic spin 
coherence can be reversibly mapped onto the collective state of the 
surrounding nuclei. The coherent transfer can be efficient and fast
and it can be used, when combined with standard resonance techniques, to 
reversibly store coherent superpositions 
on the time scale of seconds. This method can also allow for ``engineering'' entangled states of nuclear ensembles and  
efficiently manipulating the stored states. 
We investigate the feasibility of this method through a detailed analysis of 
the coherence properties of the system.
\end{abstract}
\maketitle

A broad effort is now underway to develop new techniques for coherently 
controlling quantum degrees of freedom in mesoscopic systems~\cite{awschalom:01}. 
These efforts are stimulated in part
by the proposals to use these systems as quantum bits in the context of 
quantum information science. 
The fast decoherence
associated with solid-state environments proves to be the main obstacle for experimental realization of such control.

Spin degrees of freedom of electrons confined in 
semiconductor quantum dots  are attractive qubit candidates 
\cite{loss:98,burkard:00}. Relatively long decoherence times are expected for such systems
and techniques for the coherent manipulation and measurement 
of electron spins are now being developed. For the latter, coupling of spin
and charge degrees of freedom is probably necessary.
Experimental measurements of the 
spin relaxation times indicate sub-MHz rates~\cite{fujisawa:01}, 
although it is not yet clear what will determine the 
ultimate coherence lifetimes. 

This Letter describes  a technique for greatly extending the lifetimes of 
electron-spin qubits in
confined structures by coherently mapping an arbitrary 
spin superposition state into the spins of proximal, polarized nuclei. This is achieved
by effective control of the spin-exchange part of 
 hyperfine contact interaction. After the transfer is completed, the 
resulting superpositions could be stored for a very long time -- up to seconds
-- and  mapped back into the electron spin degrees of freedom on demand.
We further show that the stored states can be manipulated using an 
extension of standard resonance techniques.

Although it is widely known that nuclear spins can possess exceptionally 
long coherence times due to their weak environmental coupling, 
single nuclear spins are very difficult to manipulate and measure in practice~\cite{kane:98}.  
In the present approach these problems are circumvented by using
collective nuclear degrees of freedom, which do not require individual 
addressing and control. We demonstrate that such collective states are extremely 
robust with respect to realistic imperfections, such as partial initial polarization and spin diffusion, and decoherence
As a result the present technique allows combines the strengths of 
electron spin (or charge) manipulation with the excellent long-term 
memory provided by nuclei. 
 
When uncontrolled, the coupling of electronic spin degrees of 
freedom to nuclei may be considered an environmental decoherence process.
Interesting features of this process arise from its non-Markovian nature~\cite{khaetskii:02,merkulov:02,khaetskii:02a}.  
The present paper shows that properly controlled coupling of 
electrons 
to nuclei can be used to greatly extend the effective coherence time of electron spins.
This study parallels recent work  
involving the use of  atomic ensembles as quantum information carriers 
\cite{duan:01}. 

To illustrate the technique, consider a single electron localized in 
quantum dot.
The effective Hamiltonian for the electron and $N$ spin-$I_0$ nuclei in a magnetic field $B_0$ along the $z$-axis is 
\begin{equation}
\hat{H} = g^* \mu_B B_0 \hat{S}_z + g_n \mu_n B_0 \sum_j \hat{I}_z^j + \hat{V}_{HF}; \label{e:ham}
\end{equation}
The first two terms of Eq.~\ref{e:ham} correspond to the Zeeman energy of the electronic and nuclear spins; the third term is the hyperfine contact interaction between the $s$-state conduction electrons and the nuclei in the dot, $\hat{V}_{HF} = \sum_j a_j \hat{\vec{I^j}} \cdot \hat{\vec{S}}$.  The coefficients $a_j = A v_0 | \psi(\vec{r}_j)|^2$ correspond to 
the one-electron hyperfine 
interaction with the nuclear spin at site $\vec{r}_j$, where $A$ is the hyperfine contact interaction constant, $v_0$ is the volume of a unit cell, and 
$\psi(\vec{r})$ is the envelope function of the localized electron.
The hyperfine term can be written $\hat{V}_{HF} = \hat{V}_D + \hat{V}_{\Omega}$ where
$\hat{V}_{D}=\sum_j a_j \hat{I}_z^j \hat{S}_z$ and 
$\hat{V}_{\Omega}  =  \sum_j a_j / 2 ( \hat{I}^j_- \hat{S}_+ +  \hat{I}^j_+ \hat{S}_-)$.
$\hat{V}_{D}$ produces an effective magnetic field for the electron
$\hat{B}_{\rm eff} = B_0 - 1/g^* \mu_B \sum_j a_j \hat{I}_z^j$, which results in the 
well-known Overhauser shift.  However, when $g^* \mu_B \hat{B}_{\rm eff} \ll \hat{V}_{\Omega}$, 
spin exchange becomes the dominate effect.

\begin{figure}
\includegraphics[width=3.4in]{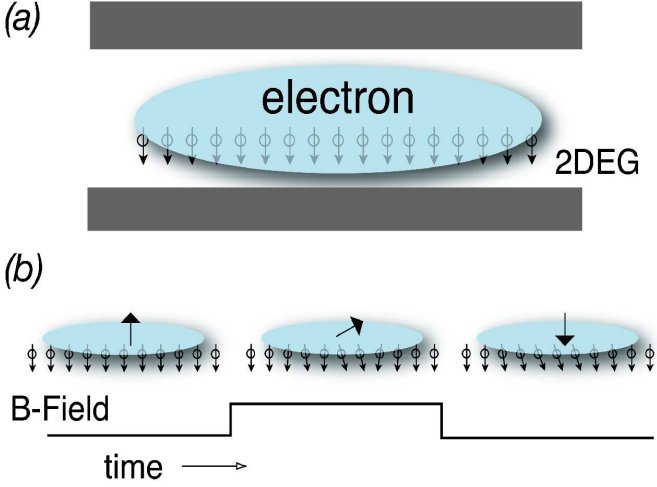}
\caption{(a) A schematic of an electron trapped in a quantum dot with polarized nuclei.
(b) Storage, where a pulse in the magnetic field starts amd brings the spins into resonance.  
After the
electron spin has flipped ($\pi$-pulse) the pulse is ended.  
Readout is the same process in reverse.
\label{f:scheme}}
\end{figure}

We start with a perfectly polarized nuclear 
ensemble $| {\bf 0} \rangle_{n} = | I_{0}, ..., I_{0}\rangle_n$, as shown in 
Fig.~\ref{f:scheme}(a).  Due to a total angular momentum conservation 
only a spin-down initial electron state can undergo non-trivial 
evolution.  When the 
dynamics are governed by $\hat{V}_\Omega$ ($\hat{B}_{\rm eff} \rightarrow 0$), there is 
a coherent exchange of excitations between electronic and nuclear degrees of freedom.
For 
the inital state $| \downn \rangle_{e} \otimes | {\bf 0} \rangle_{n}$,
spin-exchange couples this state to
the collective nuclear excitation with one flipped spin $| \upp \rangle_{e} \otimes | {\bf 1} \rangle_{n}$ with 
\begin{eqnarray}
| {\bf 1} \rangle_{n} \equiv (\sum_{j} |a_{j}|^{2})^{-1/2}
 \sum_{j} a_{j} | I_{0}, ..., (I_{0}-1)_{j}, ..., I_{0} \rangle_{n}.  \nonumber
\end{eqnarray}
Hence the evolution is given by  the two-level dynamics 
\begin{eqnarray}
\left[\matrix{ | \downn \rangle | {\bf 0} \rangle
\cr | \upp \rangle  | {\bf 1} \rangle}\right](t) =
\left[\matrix{\cos(\Omega t /2) & - i \sin(\Omega t /2) \cr i 
\sin(\Omega t /2) & \cos(\Omega t /2)}
\right] \left[\matrix{ | \downn \rangle | {\bf 0} \rangle
\cr | \upp \rangle  | {\bf 1} \rangle}\right](0), \nonumber
\end{eqnarray}
The ensemble 
displays Rabi-oscillations with an effective Rabi rate: 
\begin{eqnarray}
\Omega=\sqrt{2 I_{0}
\sum_{j} |a_{j}|^{2}}. \nonumber
\end{eqnarray}
At the same time, the spin-up electronic state 
$| \upp \rangle_e \otimes | {\bf 0} \rangle_n$ is an eigenstate of both 
$\hat{V}_D$ and $V_{\Omega}$. Hence, 
by pulsing the applied field to go from $g^* \mu_B \hat{B}_{\rm eff} \gg \hat{V}_{\Omega}$ to $\hat{B}_{\rm eff} \sim 0$ for a 
time $t = \pi / \Omega$ as diagramed in Fig.~\ref{f:scheme}(b), an 
arbitrary superposition of electronic state will undergo the following
evolution:
\begin{equation}
( \alpha | \upp \rangle_{e} + \beta | \downn \rangle_{e}) \otimes | {\bf 0} \rangle_{n} \rightarrow | \upp \rangle_{e} \otimes (\alpha | {\bf 0} \rangle_{n} + i \beta | {\bf 1} \rangle_{n}), \nonumber
\end{equation}
demonstrating that an electronic spin state can be coherently mapped into nuclei. 

The Rabi-flopping process can be controlled by removing the 
electron from the dot or by changing the effective 
magnetic field, $B_{0}$.  Away from the resonance condition 
$(|B_{\rm eff}| \gg \Omega / g^{*} \mu_B)$, the system is far detuned 
and no evolution will occur.  
For the perfectly polarized state the effective detuning  is $\delta = ( g^{*} \mu_B - g_{n} \mu_n ) B_{0} + I_{0} A + (I_{0}-1) A / N$.  The Rabi frequency 
depends upon the distribution of the $a_{j}$ coefficients, 
$\Omega = \sqrt{2 I_{0}} A / \sqrt{N} (1 + \bar{ (\Delta a^{2})}  / \bar{ a}^{2})^{1/2}$ with bars denoting
averages over the set $\{a_{j}\}$.
For a GaAs dot with $10^{4} $ nuclei and $\bar{\Delta a^{2}} \sim \bar{a}^{2}$, $A \approx 90 \mu$eV and $I_{0} = 3/2$, and
the speed of transfer is determined by
$\Omega / 2 \pi \simeq 0.6$ GHz, which exceeds the expected spin coherence time
by three orders of magnitude. 
The resonance condition is fulfilled for $| B_{\rm eff}| \ll 50$mT.
Retrieval of the stored qubit can be implemented by reversing this process: we either inject a spin-polarized electron into the dot or change the effective magnetic field, bringing the levels into resonance, and Rabi oscillations pick up at the same point as before.  

Before proceeding with a detailed description of the coherence properties
and imperfections we note that the above results can be easily generalized 
to the preparation of complex collective nuclear states.  For example, 
injection of a series of spin-down electrons into spin-up polarized nuclei 
will lead to a progressive increase of the nuclear spin.  In the basis of 
collective angular momentum, $\hat{\vec{I}} = \sum_i \hat{\vec{I^i}}$, we define
the total angular momentum nuclear states $| {\bf m} \rangle_n = | I=N I_0, I_z =  I-m \rangle_n$.  
Each electron can effect the transfer $| \downn \rangle_e \otimes | {\bf m} 
\rangle_n \rightarrow | \upp \rangle_e \otimes | {\bf m+1} \rangle_n$.  When 
injected electrons are prepared 
in different superposition states this process can be used to effectively 
``engineer''
collective states of nuclear ensembles. In fact using a proper sequence of 
electrons an arbitrary state of the type $| {\bf \Psi}\rangle_n = \sum_{m=0}^{I} c_m 
| {\bf m} \rangle_n$ can be prepared~\cite{foot:01}. We note in particular that the 
highly entangled states of the kind $( |{\bf 0} \rangle_n +|{\bf m}\rangle_n) /\sqrt{2} $, with 
large ${\bf m}$ could be used for high-resolution NMR spectroscopy in analogy 
with
related atomic physics studies~\cite{bollinger,huelga}.  Such states
can also be prepared by manipulating one electron in the dot with 
fast electron-spin 
resonance (ESR) pulses.

Injection of polarized electrons, combined with ESR pulses, 
can also be used to perform manipulation of the 
stored nuclear state.  For example, the qubit stored in nuclear spin could be 
mapped
into the injected electron and an ESR pulse can be used to rotate it. Subsequently, 
it can be mapped back to nuclei. Alternatively virtual, off-resonant 
($|\delta| \gg \Omega$) coupling of  storage states to electron spin can be 
used to shift the resonance frequencies of transitions 
$| {\bf m} \rangle_n \rightarrow   | {\bf m+1} \rangle_n$ relative 
to each other. For example in the case of two lowest states $m=1,2$ the 
differential shift is on the order of  $ \Omega^4 /  |\delta|^3 $. 
Whenever
this shift is large compared to decoherence rate and the spectral width of
excitation pulse, the lowest two states of collective manifold 
$| {\bf 0} \rangle_n, | {\bf 1} \rangle_n$ can be considered
as an effective two level system  and can be manipulated through NMR 
pulses and other means. 
These ideas could be extended to the spin-exchange coupled qubits proposal ~\cite{loss:98}.

We now turn to the consideration of various decoherence mechanisms and 
imperfections that limit the performance of the storage technique.   
To evaluate the effects of spatial inhomogeneity we note that 
the collective state $| \upp \rangle_e \otimes | 1 \rangle_n$ is not an eigenstate of $\hat{V}_D$ unless the $a_j$'s are identical. The effect is determined by 
the distribution of eigenenergies under $\hat{V}_D$ of the states $| \upp \rangle_e \otimes | I_0, ..., I_0 - 1_j, ... , I_0 \rangle_n$, given by
\begin{eqnarray}
E_j = ( g^{*} \mu_B / 2 - g_{N} \mu_N I_0 ) B_{0} + I_{0} A / 2 - a_j / 2 
\nonumber
\end{eqnarray}
Since $\bar{ \Delta E^2} \sim \bar{ \Delta a^2}$, the inhomogeneous linewidth 
is $\sim \bar{a} = A/N$.  Hence inhomogeneous broadening is smaller than the relevant time scale for Rabi-flopping by a factor of $\sqrt{N}$, and is negligible during transfer operation.
After the mapping, its effect can be mitigated by either removing the electron from the dot, thereby turning off the hyperfine interaction entirely, or by using ESR spin-echo techniques to reverse the phase evolution~\cite{spin-echo}. 

The leading decoherence process for the stored state is 
nuclear-spin diffusion, with dephasing rates in the kHz domain.
However, techniques from NMR can be used to 
mitigate this effect~\cite{mehring,paget}.  
By applying fast NMR 
pulse sequences~\cite{whh} to rotate the nuclear spins 
the time average of the leading terms in 
the dipole-dipole Hamiltonian can be reduced to zero, leaving only residual dephasing due to second-order dipolar 
effects, pulse imperfections and
terms due to the finite length of the averaging sequence.  
These phenomena have been studed for several decades for solid-state NMR 
systems and well-developed techniques routinely reduce $T_2$ by several orders of magnitude~\cite{mehring}, down to sub-Hz levels.  Hence, coherent qubit storage on the time scale of seconds seems feasible.

To evaluate the effects of partial polarization on the storage fidelity, we use 
the Heisenberg picture. In the homogeneous case $(a_i = a = A/N)$, 
the Dicke-like collective operators defined above yield 
$\hat{I}^2$ and $\hat{J}_z = \hat{S}_z + \hat{I}_z$ as the constants of motion.  We 
consider operator equations of motion $\dot{\hat{A}} = i [ \hat{A},\hat{H}]$, for the three operators $\hat{S}_{z}, \hat{S}_{+} \hat{I}_{-}$, and $\hat{S}_{-} \hat{I}_{+}$, which commute with the constants of motion.  
We replace $\hat{I}_{z}$ terms in the resulting equations with $\hat{J}_{z} - \hat{S}_{z}$ and use the identity $\hat{I}_{+} \hat{I}_{-} = \hat{I}^{2} - \hat{J}_{z}^{2} - [\hat{I}_{-},\hat{I}_{+}]/2$ to put the equations in terms of 
constants of motion and the three operators we 
look to solve:
\begin{eqnarray}
\frac{d}{dt} \hat{S}_z & = & a  \frac{\hat{S}_+ \hat{I}_- - (\hat{S}_+ \hat{I}_-)^{\dag}}{2 i}  \\ 
\frac{d}{dt} (\hat{S}_+ \hat{I}_-) & = & i  [  (g^* \mu_B - g_n \mu_n) B_0  + a (\hat{J}_z - 1)] (\hat{S}_+ \hat{I}_-) \nonumber \\
 & - & i a  ( \hat{I^2} - \hat{J}_z^2 + 1/4) \hat{S}_z.  
\end{eqnarray}
It is convenient to choose new constants of motion, $\hat{\delta} =   (  (g^{*} \mu_B - g_{n} \mu_n ) B_0  + a (\hat{J}_z - 1))$ and $\hat{\Omega} = a (\hat{I^2} - \hat{J}_z^2 + 1/4)^{1/2}$.  These commute with each other and with $\hat{S}_z, \hat{S}_+ \hat{I}_-$.  As these equations are similar to those for two-level atoms in a field, we make the Bloch vector identifications:
\begin{eqnarray}
\hat{U} & = & -a / \hat{\Omega} \frac{ \hat{S}_+ \hat{I}_- + (\hat{S}_+ \hat{I}_-) ^{\dag}}{2} , \nonumber \\
\hat{V} & = & -a / \hat{\Omega} \frac{ \hat{S}_+ \hat{I}_- - (\hat{S}_+ \hat{I}_-) ^{\dag}}{2 i} , \nonumber \\
\hat{W} & = & \hat{S}_z.
\end{eqnarray}
and Eqns. 3--4 become $\dot{\hat{\vec{M}}} = \hat{\vec{M}} \times \hat{\vec{\omega}}$ with $\hat{\vec{\omega}} = ( \hat{\Omega}, 0, -\hat{\delta})$.  
The Bloch vector operator, $\hat{\vec{M}}$, will rotate about the axis defined by $\hat{\vec{\omega}}$ at a frequency $\hat{\omega}_0 = \sqrt{ \hat{\delta}^2 + \hat{\Omega}^2}$.  For no initial electron spin-nuclear spin correlation, we can easily solve for $\hat{S}_z$ and find
\begin{equation}
\langle \hat{S}_z (t) \rangle = \langle \frac{ \hat{\delta}^2 + \hat{\Omega}^2 \cos ( \hat{\omega}_0 t)}
{\hat{\omega}_0^2} S_z (0) \rangle. \label{e:sz}
\end{equation} 

For the perfectly polarized nuclear state, Eq.~\ref{e:sz} gives $\langle \hat{S}_z \rangle = -1/2 \cos ( A \sqrt{2 I_0 / N} t)$ for the spin-down species and $1/2$ for the spin-up species, exactly replicating the fully polarized behavior.  In the case of partial polarization, even though we will set $\langle \hat{\delta} \rangle = 0$, $\langle \hat{\delta}^2 \rangle$ remains finite.  The first part of Eq.~\ref{e:sz} prevents complete transfer, to order $\langle \hat{\delta}^2 \rangle / \langle \hat{\omega}_0^2 \rangle$.  For partial polarization $P < 1$, $\langle \hat{\omega}_0^2 \rangle \approx a^2 N (2 I_0 + O(1-P))$.  
In a thermal state, all 2-particle expectation values factor, so $\langle \hat{\delta} \rangle = 0$ and
\begin{eqnarray}
\langle \hat{\delta}^2 \rangle = a^2 ( \langle \hat{J}_z^2 \rangle - \langle \hat{J}_z \rangle^2 ) \approx N a^2 (1 - P ) 2 I_0 .
\end{eqnarray}
Accordingly, the error scales as
\begin{eqnarray}
\frac{\langle \hat{\delta}^2 \rangle}{\langle \hat{\omega}_0^2 \rangle} \approx 1 - P .
\end{eqnarray}
Remarkably, this result demonstrates that efficient transfer is possible even with many nuclei in the ``wrong'' state as long as the average polarization per nuclei is high.

\begin{figure}
\includegraphics[width=3.4in]{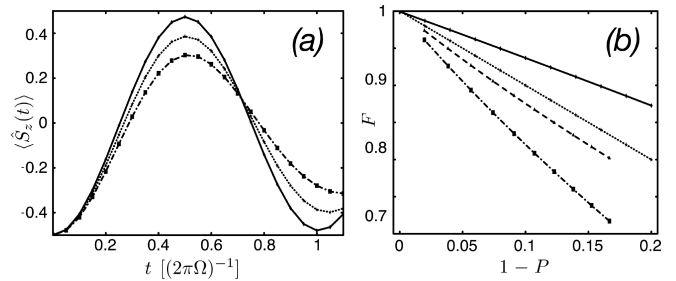}
\caption{Transfer characteristics for different nuclear polarizations.
{\em (a)} The evolution of $\langle S_z(t) \rangle$ for three nuclear polarizations, $P=0.96$ (solid curve), 0.82 (dotted), and 0.67 (dot-dash) versus time in units of the $P=1$ Rabi time.
{\em (b)} Solid line: Storage fidelity $F$ for inhomogeneous effects ($N=10^4$) versus error in nuclear polarization, $1-P$.  Dotted line: analytical estimate of thermal effects.  Dashed line: numerical simulations of thermal effects.  Dot-dash line: product of inhomogeneous and thermal results.
\label{f:errorestimate}
}
\end{figure}

We modeled this system numerically with an initial thermal nuclear state.
Oscillation of $\langle \hat{S}_z \rangle$ for several polarizations is shown in Fig.~\ref{f:errorestimate}(a).
The effect of partial polarization is immediately apparent; the transfer peak is less than 1/2, and the Rabi frequency decreases.  The difference between the initial electron state and the state after once cycle of storage and retreival corresponds to the storage fidelity $F$.  Fig.~\ref{f:errorestimate}(b) shows that the analytical estimate is close to the numerically calculated values; thus, the transfer error can, to a large extent, be explained by the thermal uncertainty in $\hat{\delta}$.  The residual effect most likely stems from phase mismatching, as measured by the broadening of $\omega_0$, such that at $t= ( \pi \langle \omega_0 \rangle )^{-1}, \langle \cos ( \omega_0 t ) \rangle = 1 - O(1-P)$.  

Inhomogeneous broadening for a thermal initial nuclear state is somewhat more pronounced than the fully polarized case first considered.  The $\hat{V}_D$ inhomogeneity causes slow dephasing of the stored state, and the inhomogeneous coupling in $\hat{V}_{\Omega}$ results in leakage in readout of the stored state into a set of states orthogonal to the original nuclear state.  The results shown in Fig.~\ref{f:errorestimate}(b) were calculated for a distribution of $a_i$'s corresponding to a gaussian $| \psi(\vec{r}) |^2$.
We plot the estimated fidelity for a complete storage and readout cycle as a function of the initial polarization of nuclear spins.  The total expected fidelity,
is approximated by the product of these two results (Fig.~\ref{f:errorestimate}(b)).  Hence, only modest nuclear polarizations are required to obtain a high fidelity of storage.

In summary, we have demonstrated that is possible to reliably map the
quantum state of a spin qubit onto long-lived collective nuclear spin states.
The resulting states have long coherence times, 
and can be retrieved on demand. 
Furthermore, the stored states can be efficiently 
manipulated and similar techniques can be used for quantum  state engineering
of collective nuclear states. 

Experimental implementation of these ideas requires preparation of nuclear spin polarizations in the vicinity of confined electrons. Optical pumping of nuclear spins has 
demonstrated polarizations in GaAs 2D electron gases on the order of 90\%~\cite{salis:01} and 65\% in self-assembled dots~\cite{gammon:01}, and forced spin-flips through quantum hall edge states~\cite{dixon:97} has a claim of a similarly high polarization ability (85\%).
We anticipate that combining either of 
these techniques with electron localization in quantum dots  would be a promising avenue of study. The methods outlined in this paper can 
be used to further increase the nuclear polarization. 
A current of spin-polarized electrons passing through the quantum dot with 
a dwell time $\tau_{\rm dwell} < \pi / \Omega_{0}$ will result 
in spin transfer, thereby increasing nuclear polarization. 
By keeping $B_{\rm eff}$ tuned to zero with increasing nuclear polarization 
the spin flip-flop remains resonant and, when combined with dephasing to prevent saturation, 
leads to efficient cooling, similar to a recent proposal~\cite{eto:02}.  

Coherence properties of the spin-exchange process
could be probed in transport measurements. 
For example, sending spin polarized currents through the 
quantum dot in which the spin-exchange interaction is tuned to resonance
will result in collapses  and revivals of the electron spin polarization
that will be a periodic function of the dwell time in the dot. Those can be 
measured using spin-filter techniques. 
For a given polarization $P$ and $\tau_{\rm dwell}$, the spin will be rotated by $\Omega(P) \tau_{\rm dwell} / \pi$.  Note that repolarization will be necessary after $\sim \sqrt{N}$ nuclear spins have been flipped.

Practical applications of the storage and retrieval techniques and 
manipulation of stored states requires time-varying
 control  over the spin-exchange coupling. This can be accomplished by 
using a pulsed magnetic field of order 50mT for a few ns,
 by engineering the electron $g$-factors ~\cite{salis:01b, matveev:00}, or
by optical AC Stark shifts~\cite{salis:01}
These techniques can be combined with a number of avenues for entanglement
and manipulation of the electronic spin and charge states currently under exploration~\cite{awschalom:01,loss:98}.  Finally, quantum memory can facilitate 
implementation and reduce  
scaling problems for more  ambitious tasks such as quantum error 
correction \cite{steane:99} or
quantum repeaters \cite{duan:01}.  

We thank J.A. Folk, Y. Yamamoto, and D. Loss for helpful discussions. This
work was supported by ARO, NSF, and by
David and Lucille Packard Foundation.

\end{document}